   \documentclass[cup6b]{cupbook}
   \usepackage{graphicx}
   \usepackage{natbib}

\begin{document}

\author
{
  R. C. Thomas                                                  \\
  University of Oklahoma, Department of Physics and Astronomy,  \\
  440 W. Brooks Street Rm. 131, Norman, Oklahoma 73071          \\
  (Present Address:
  Lawrence Berkeley National Lab,                               \\
  1 Cyclotron Road MS 50R5008, Berkeley, California 94720)
}

\chapter{ Synthetic Spectrum Methods for Three-Dimensional Supernova Models }

{\it
Current observations stimulate the production of fully three-dimensional
explosion models, which in turn motivates three-dimensional spectrum synthesis
for supernova atmospheres.  We briefly discuss techniques adapted to address
the latter problem, and consider some fundamentals of line formation in
supernovae without recourse to spherical symmetry.  Direct and detailed
extensions of the technique are discussed, and future work is outlined.
}

\section{ Introduction }

Spectrum synthesis is the acid test of supernova modelling.  Unless
synthetic spectra calculated from a hydrodynamical stellar explosion
model agree with observations, the model is not descriptive.  Some
explosion modellers contend that only three-dimensional (3D) models
faithfully describe the physics of the real events.  If this is so,
then the evaluation of those models requires solutions to the 3D model
supernova atmosphere problem.  These solutions require full {\it
detail}, the inclusion of as much radiation transfer physics as possible.
Otherwise, a bad fit of a synthetic spectrum to an observed one might
have less to do with the accuracy of the hydrodynamical model, and
more to do with the shortcomings of the radiation transfer procedure.

On the other hand, solutions (of a sort) to the ill-posed inverse
problem constrain parameter space available to hydrodynamical models.
Fast, iterative, parameterized fits to observed spectra characterize
the ejection velocities and identities of species found in the line
forming region.  Most importantly, the procedure reveals species that
{\it cannot} be identified by simply Doppler-shifting line lists on
top of observed spectra in search of feature coincidences.
Generalizing this {\it direct} analysis technique to 3D is key to
constraining the geometries of real explosions.

This proceedings contribution briefly describes some steps toward the
complimentary goals of detailed and direct analysis in 3D, with an emphasis on
pedagogy.  For an in-depth application of the more detailed technique, refer to
the contribution of D. Kasen.

\section{ Approach }

Work underway to extend spherically symmetric non-LTE modelling codes to 3D
could take the better part of a decade.  An alternative approach, which yields
a direct analysis code along the way, is to begin with a simple 3D code and
augment its physics details.  The ultimate goal is complete non-LTE radiation
transfer in 3D with full and realistic treatment of the boundary conditions at
depth.  This means that the evolution of radiation from deposition to escape is
modelled without the central ``light bulb'' approach.

We embark on the journey toward full 3D non-LTE modelling from the elegant and
humble shores of the Sobolev method \citep{Castor1970, Rybicki1978}.  
The Sobolev method greatly reduces
the scale of the wavelength-domain of the problem by approximating line 
transfer, but does so at an accuracy cost.  When faced with the alternative,
some inaccuracy in the line transfer is acceptable.  Full solutions to the 3D
radiation transfer problem are simply not possible yet, so the limited
inaccuracies (and the awareness of them) seems a small price to pay for
progress.

The particular implementation of the Sobolev method is the Monte Carlo
technique, based largely on the formalism described primarily by L. Lucy in a
series of papers \citep[e.g.,][]{Lucy1999}.
The key innovation described in those papers we call
the equal-energy packet (EEP) technique.  In the EEP picture, individual photon
trajectories are not simulated, but rather monochromatic photon packets
of equal energy
propagate through the model atmosphere.  This paradigm obviates recursive
trajectory calculations which make development and extension of transfer codes
difficult.  Of course, the scalability of the Monte Carlo technique is one of
its greatest virtues.  Provided the model atmosphere need not be split up
across distributed nodes, communication proves almost nonexistent.
Most importantly, this implementation micro-manages the energy
conservation (flux divergence-less-ness) of the radiation field at all positions
(or depths).  Scattering is coherent in the comoving frame.  Absorption and
re-emission is accomplished through roulette-wheel selection (producing either
a simple equivalent two-level atom or non-LTE source function).  The Monte
Carlo algorithm propagates variations in the radiation field instantaneously
(provided enough packets are used), and this makes a simple 
$\Lambda$-iteration work.

Clearly, waiting for enough packets to exit the model atmosphere in any given
direction of interest is not the most efficient use of limited computer time.
Hence, at least for flux spectra, we can easily use the packets to establish
the radiation field (energy density) throughout the envelope.  From that, we
derive the source function at all points and compute the emergent spectrum for
any line of sight, a kind of ``formal integral,'' again simplified by the
Sobolev approximation.  This technique greatly reduces the number of packets
required to build a spectrum.

Other extensions to the method besides the formal integral for emergent spectra
include polarized transfer and a way of treating energy deposition from
radioactive decay in the ejecta.  The propagation of Monte Carlo Stokes vectors 
for computing polarization spectra and progress away from the central ``light
bulb'' approximation are exemplified in D. Kasen's contribution.

\section{ Simple Constraints on Nonsphericity }

The starting question is a rather obvious one, but its answer is fundamental.
If deviations from spherical symmetry occur in supernova atmospheres, what is
the corresponding detection threshold for their evidence to appear in flux
spectra?  The changes to line profiles resulting from nonsphericity are easy to
understand, but we seek means of quantitatively exploiting the results to
constrain supernova geometry.

\begin{figure}
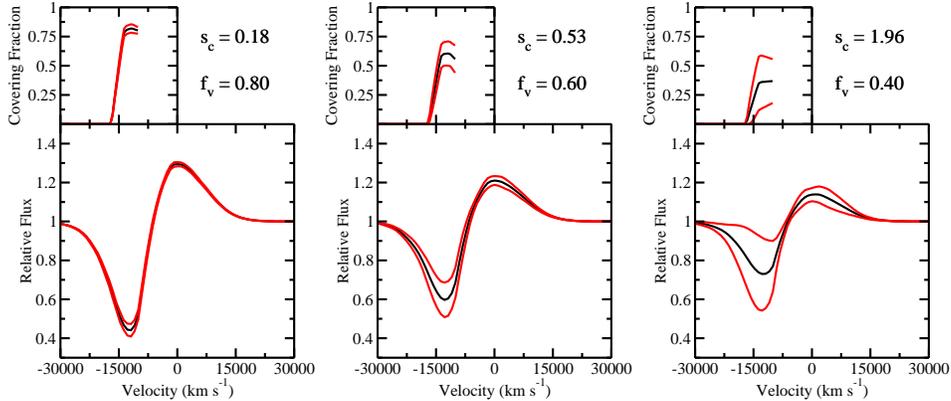

 \centering
 \includegraphics[scale=0.29,clip]{noise.cs=001.ff=080.eps}
 \includegraphics[scale=0.29,clip]{noise.cs=004.ff=060.eps} 
 \includegraphics[scale=0.29,clip]{noise.cs=016.ff=040.eps}
 \caption
 {
  Sample results from a simple experiment.  
  The parameter $f_V$ is the fraction of the line-forming region with non-zero
  optical depth, and $s_c$ is the ratio of the cubical ``clump'' edge length
  to the photospheric radius.  Average line and covering fraction profiles 
  (dark curve) plus ``1-$\sigma$'' deviation (light curves) from the average 
  profiles are shown.
 }
 \label{fig:clumpExamples}
\end{figure}

To find out, we conduct a very simple experiment.  We generate a series of
distributions of Sobolev optical depth covering a range of nonsphericity
scales.  For each model, we compute line profiles from a large number of lines
of sight.  This gives us a sense of what sort of diversity can be generated by
what size of perturbation.  In Figure \ref{fig:clumpExamples} are line and
covering fraction profiles averaged over 100 lines of sight.  The scatter 
around the average is summarized by a 1-$\sigma$ deviation from the average.
The covering fraction for a given velocity is defined as the fraction of the 
projected photosphere surface obscured by Sobolev optical depth exceeding 1.
We deduce three facts from the results.  First, nonsphericity of the
kind considered here can most strongly influence absorption features.  Second,
these nonsphericities only weakly influence emission features.  (Both of these
previous points are reversed if the line source function significantly exceeds
that of resonance scattering).

Third and most importantly, the diversity trend as a function of the line of
sight is directly correlated with the photospheric covering fraction.  Hence,
it is possible to suggest a threshold scale of clumpiness below which clumping
goes undetected in flux spectra.  At first this result seems purely academic,
since nature presents us with only one line of sight to a given supernova.

However, we can apply the result to make a claim about models for
spectroscopically normal Type Ia supernovae.  Such supernovae are
spectroscopically homogeneous, and we find that the depth of the Si II feature
in these events is a fairly repeatable 0.7 times the local continuum.  In fact,
the measured scatter seems to suggest that perturbations like those explored
here must be smaller than 10\% the size of the photosphere area if they are
indeed present.  More importantly, if pure 3D deflagrations exhibit such large
scale perturbations leading to wildly fluctuating line covering fractions as a
function of perspective, they cannot account for such events.

\section{ Direct Analysis in 3D }

The oft-repeated goals of spherically symmetric direct analysis are to identify
lines and velocity intervals within the line forming region where the parent
ions of those lines are found.  In direct analysis, special attention is given
to treating line blending, since this is an important feature of supernova
spectra.  Without this attention, we have seen that identifications are
problematic, and these problems make it more difficult to narrow down the range
of hydrodynamical models that are worth pursuing.

To provide the same direct analysis capabilities, but without recourse to
spherical symmetry, we developed a code called \texttt{Brute}, based on the
earlier, spherically symmetric (and non-Monte Carlo) code \texttt{Synow}.  The
basic picture is the same, except that instead of radial functions of Sobolev
optical depth in a reference line of each ion included, we use a template
(constructed in any fashion) that need not be spherically symmetric.  

An application of this code to the unique Type Ia supernova 2000cx appears in
\citet{Thomas2003}.  The maximum light spectra of this object exhibited
unusual, narrow high-velocity features, particularly in the Ca II infrared
triplet.  Using alternate 1D and 3D models of optical depth in Ca II, we find
that partial blocking of the photosphere by a nonspherical ejecta distribution
can help explain some of these features.  Simultaneous fits are attempted to
the corresponding Ca II UV feature, and we find that the 3D distribution is
less problematic than the 1D one.  The origin of the high-velocity material in
this supernova is still a mystery, but how much of the observed diversity in
these objects is due to 3D distributions of ejecta at high velocity?  More
modelling of more objects is needed.

\begin{figure}
 \centering
 \includegraphics[scale=0.4000,clip]{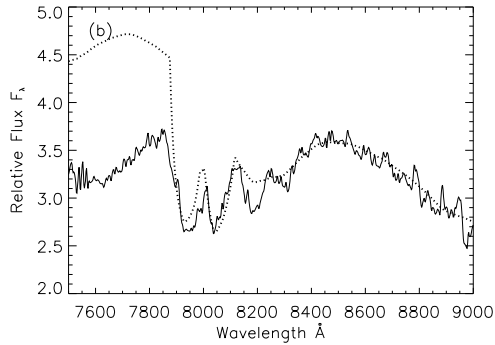}
 \caption
 {
  High velocity Ca II fit using a 3D optical depth parameterization.
  From \citet{Thomas2003}.
 }
 \label{fig:sn00cx}
\end{figure}

\section{ Detailed Analysis in 3D }

A special characteristic of the Monte Carlo technique is that it allows for
simple solution to the radiative equilibrium problem.  This permits more
self-consistent modelling of emergent spectra from real hydrodynamical models.
Given a composition and density structure (in any geometry), we use the
$\Lambda$-iteration procedure to construct self-consistent temperature
structures.

\begin{figure}
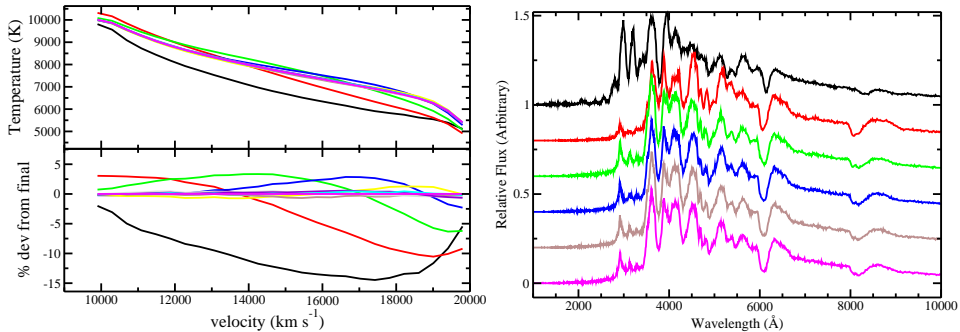

 \centering
 \includegraphics[scale=0.2513,clip]{w7noe.temp.eps}
 \includegraphics[scale=0.2513,clip]{w7noe.spec.eps}
 \caption
 {
  Test results for a mixed W7 model.  Rapid convergence in the temperature
  structure is evident.  Resulting spectra from iterations 1, 3, 5, 7, 9 and 
  10 are shown to the right.
 }
 \label{fig:lte}
\end{figure}

In Figure \ref{fig:lte}, temperature structures and spectra are shown from
10 iterations to a converged model.  Though this particular model is
spherically symmetric (a W7-like model mixed above 9000 km s$^{-1}$), 
the convergence speed is striking.  This provides us with
hope that (at least for now) we can begin modelling real hydrodynamical
explosion models, at least in LTE.

\section{ Conclusion }

The path to fully detailed 3D non-LTE spectrum synthesis for supernova models
is clear.  Including non-LTE and continuum transfer effects will permit us to
examine core collapse supernovae more closely, to help unlock the connection
between supernovae and gamma-ray bursts.  The eventual goal of dispensing with
the Sobolev approximation will also be reached, and progress is already
underway with realistic lower boundary conditions and gamma-ray transport.  

Eventually, 3D hydrodynamical stellar explosion models will be carried to
the homologous expansion phase.  Those models will be converted into flux and
polarization spectra to be compared with observations.  Until then, there is
much computer time to be burned.

This work is supported by grant HST-AR-09544-01.A, provided by NASA through
the STScI, operated by the AURA, Inc., under NASA contract NAS5-26555.

\end{document}